\documentclass{isprs} % isprs class modified 23-04-2019 (Dennis Wittich)
\usepackage{setspace}
\usepackage{graphicx,subfigure}
\usepackage{mwe}
\usepackage{diagbox}
\usepackage{geometry} % added 27-02-2014 Markus Englich
\usepackage{epstopdf}
\usepackage[T1]{fontenc}
\usepackage{floatrow}
\usepackage[british]{babel} 
\usepackage[hang]{footmisc}
\usepackage{multirow}
\usepackage[utf8]{inputenc}
\usepackage{amsmath,amssymb}
\usepackage{xcolor}

\usepackage{float}
\floatstyle{plaintop}
\restylefloat{table}
\usepackage[]{natbib} 

\begin{document}
\title{Lake Ice Detection from Sentinel-1 SAR with Deep Learning}
% KAO: Remove extra spacing
\author{
 Manu Tom\textsuperscript{1}\thanks{equal contribution}~~\thanks{corresponding author},~~Roberto Aguilar\textsuperscript{2}\footnotemark[1],~~Pascal Imhof\textsuperscript{1}, Silvan Leinss\textsuperscript{3}, Emmanuel Baltsavias\textsuperscript{1}, Konrad Schindler\textsuperscript{1}}
% KAO: Remove extra newline
\address{
	\textsuperscript{1} Photogrammetry and Remote Sensing Group, ETH Zurich, Switzerland, \textsuperscript{2} Space Center, Skoltech, Russia\\
	\textsuperscript{3} Earth Observation and Remote Sensing Group, ETH Zurich, Switzerland, leinss@ifu.baug.ethz.ch\\(manu.tom, manos, schindler)@geod.baug.ethz.ch, paimhof@student.ethz.ch,	roberto.aguilar@skoltech.ru
}
% If the corresponding author is NOT the final author, always add a % space before the subsequent comma, i.e.
% first author name\textsuperscript{a,}\thanks{Corresponding author} , % second author name \textsuperscript{b}, etc.
% thanks to Niclas Borlin 05-05-2016
\commission{III, }{} %This field is optional.
\workinggroup{III/9} %This field is optional.
\icwg{}   %This field is optional.
% KAO: Use times symbol
\abstract{\textit{Lake ice}, as part of the Essential Climate Variable
  (ECV) \textit{lakes}, is an important indicator to monitor climate
  change and global warming. The spatio-temporal extent of lake ice
  cover, along with the timings of key phenological events such as
  \textit{freeze-up} and \textit{break-up}, provide important cues
  about the local and global climate. We present a lake ice monitoring
  system based on the automatic analysis of Sentinel-1 Synthetic
  Aperture Radar (SAR) data with a deep neural network. In previous
  studies that used optical satellite imagery for lake ice monitoring,
  frequent cloud cover was a main limiting factor, which we overcome
  thanks to the ability of microwave sensors to penetrate clouds and
  observe the lakes regardless of the weather and illumination
  conditions. We cast ice detection as a two class (\textit{frozen,
    non-frozen}) semantic segmentation problem and solve it using a
  state-of-the-art deep convolutional network (CNN). We report results
  on two winters ($2016$-$17$ and $2017$-$18$) and three alpine lakes in
  Switzerland. The proposed model reaches mean Intersection-over-Union (mIoU) scores \textgreater90\% on average, and \textgreater84\% even for the most difficult lake. Additionally, we perform cross-validation tests and show that our algorithm generalises well across unseen lakes and winters.}

\maketitle
%\saythanks % added 28-02-2014 Markus Englich
\section{INTRODUCTION}
Climate change is one of the main challenges humanity is facing today,
calling for new methods to quantify and monitor the rapid change in
global and local climatic conditions.
Various lake observables are related to those conditions and provide
an opportunity for long-term monitoring, among them the duration and
extent of lake ice. Remote sensing of lake ice also fits well with the
Climate Change Initiative (CCI+, 2017) by the European Space Agency
(ESA), where \textit{lakes} and \textit{lake ice} were newly
included. Additionally, CCI+ promotes long-term trend studies and
climate studies, as recognised by the Global Climate Observing System
(GCOS). Furthermore, lake ice influences various economic and
social activities, such as winter sports and tourism, hydroelectric
power, fishing, transportation, and public safety (e.g., winter and
spring flooding due to ice jams). In addition, its impact on the
regional environment and ecological systems is significant, which
further underlines the need for detailed monitoring.
%
%\begin{figure}[]
%   \centering
%    \includegraphics[width=0.49\textwidth]{figures/3lakes_comp.png}\\
    %\subfigure[Skating tracks on the during winter season caught on a webcam image.]{\includegraphics[width=0.3\textwidth]{figures/patterns_st_moritz.jpg}}
    %\caption{Sentinel-1 SAR data composite (\textit{VV, VH}) of lake St.~Moritz showing the lake in three different states: water (left), snow with skating tracks (middle), and break-up (right).} 
%\label{fig:title_figure}
%\end{figure}

Satellites are a secure source for remote sensing of the cryosphere
and for sustainable, reliable, and long term trend analysis. Additionally, satellite images are currently the only means to monitor large regions systematically and with short update
cycles. This increasing importance of satellite observations has also
been recognised by the GCOS. Recently, \cite{tom_lakeice_2018}
proposed a machine learning-based semantic segmentation approach for
lake ice detection using low spatial-resolution ($250$m-$1000$m)
optical satellite data (MODIS and VIIRS). Although the nominal temporal resolution of those sensors is very good (daily coverage), the main drawback of this methodology is frequent data loss due to clouds, which reduces the effective temporal resolution. This is critical, since important phenological variables depend on frequent and reliable observation. In particular, the \textit{ice-on} date is defined as the first day when the lake surface is (almost) completely frozen and remains frozen on the next day, and \textit{ice-off} is defined symmetrically as the first day where a significant amount of the surface is liquid water, and remains in that state for another day (\citeauthor{franssen_scherrer2008} and Scherrer, 2008). The GCOS accuracy requirement for these two dates is $\pm 2$
days. Systems based on optical satellite data will fail to determine
these key events if they coincide with a cloudy period.
Moreover, low spatial resolution of MODIS and VIIRS is also a
bottleneck for spatially explicit ice mapping. Higher resolution
optical sensors like Landsat-8 or Sentinel-2 do not provide a
solution, due to their low temporal resolution and susceptibility to
clouds.
On the contrary, Sentinel-1 represents a favourable trade-off between
spatial and temporal resolution. Additionally, Radar is unaffected by
clouds, which in many regions is a considerable advantage.
\begin{figure}[!ht]
  \centering
    \subfigure[Non-frozen (01.09.2016)]{\includegraphics[width=0.4\textwidth]{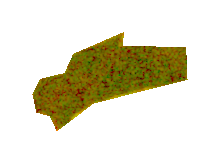}}
    \subfigure[Freeze-up (10.01.2017)]{\includegraphics[width=0.4\textwidth]{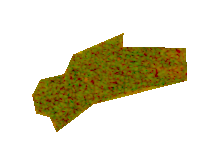}}\\
    \vspace{-0.5em}
    \subfigure[Snow with skate tracks (08.02.2017)]{\includegraphics[width=0.4\textwidth]{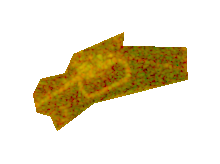}}
    \subfigure[Break-up (23.03.2017)]{\includegraphics[width=0.4\textwidth]{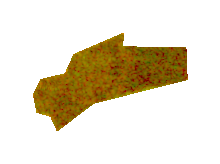}}
    %\hfill
    \vspace{-1.2em}
    \caption{Examples of RGB composites of Sentinel-1 SAR data (RGB = [\textit{VV, VH, 0}]) of lake St.~Moritz showing the lake in the four different states specified in the sub-captions.} 
\label{fig:title_figure}
\end{figure}

Here we propose to use Sentinel-1 SAR data, which mostly meets the
requirements of lake ice monitoring, and additionally comes for free
and with a commitment to ensure continuity of the observations. Its
spatial and temporal resolution (GSD ca.\ 10 m / revisit period if 1.5
days) make it possible to derive high-resolution ice maps almost on a
daily basis.
For completeness, we mention that, taking into account estimation
uncertainty, the temporal resolution of Sentinel-1 falls just short of the 1-day
temporal resolution requirement of lake ice monitoring, it can still provide an excellent ``observation
backbone'' for an operational system that could fill the gaps with
optical satellite data \citep{tom_lakeice_2018} or webcams
\citep{muyan_lakeice_2018}.

Converting a Sentinel-1 image to a lake ice map boils down to 2-class
semantic segmentation, i.e., assigning each lake pixel to one of two
classes, \textit{frozen} or \textit{non-frozen}. We do this with the
\textit{Deeplab v3+} semantic segmentation network
\citep{deeplabv3plus2018}. Examples of Sentinel-1 SAR composites over
the lake St.~Moritz is visualised in Fig.~\ref{fig:title_figure}, showing the \textit{VV} amplitude in the red channel, and the \textit{VH} amplitude in the green channel. The examples include the states \textit{non-frozen} (01.09.2016, water), \textit{freeze-up}
(10.01.2017), \textit{frozen} (08.02.2017, snow on top of ice) and a
\textit{break-up} date (23.03.2017).

%In general, it is very difficult to make a visual interpretation of the SAR data. However, the skating tracks that are clearly visible in a webcam image (see Fig. \ref{fig:title_figure}b) can also be seen in Fig. \ref{fig:title_figure}b (middle).

\textbf{Target lakes and winters.} We analyse three selected lakes in
Switzerland (\textit{Sils, Silvaplana, St.~Moritz}, see
Table~\ref{table:lakes}) over the period of two winters ($2016$-$17$ and
$2017$-$18$). These three lakes are located close to each other in the
same geographic region, referred to as \textit{Region Sils}. The lakes
are comparatively small and situated in an Alpine environment, and
they are known to reliably freeze over completely every year during
the winter months.  For the two winters $2016$-$17$ and $2017$-$18$, all available images were collected for the nine months between September $1$ and May $31$.
%\par
After back-projecting the digitised lake outline polygons from Open Street Map
(OSM) on to the SAR images, for each lake, we extract the \textit{lake pixels} which lie inside the lake outline.
%divide the lake pixels into two categories (a) \textit{clean pixels} which lie completely inside the lake and (b) \textit{mixed pixels} which lie partially on the lake.
In low-spatial resolution satellite images such as \textit{MODIS} and
\textit{VIIRS}, only few such \textit{lake pixels} are available
\citep{tom_lakeice_2018}, which makes the analysis of very small lakes such as St.~Moritz difficult or even impossible. Thanks to the higher spatial resolution, the Sentinel-1 time series provides us with millions of lake pixels, which makes it possible to train powerful deep learning models for segmentation, which are extremely data-hungry.

\textbf{Contributions.}
We address the problem of lake ice detection from Sentinel-1 SAR data,
as an alternative to optical satellite data which is impaired by
clouds. In the process, we show that a deep learning model pre-trained
on an optical RGB dataset can nevertheless be re-used successfully as
initialisation for fine-tuning to Radar data. To our knowledge, our
work is the first one that utilises Radar data and deep learning for
lake ice detection, though it has been used for sea ice analysis.
\begin{table}[ht]
\small
% 	\vspace{-0.2cm}
	    \centering
	    \vspace{-0.5em}
	    \begin{tabular}{ |cccc| } 
		\hline
		                   &   \textbf{Sils} & \textbf{Silvaplana} & \textbf{St.~Moritz}  \\ 
		\hline
		Area ($km^{2}$)   & $4.1$ & $2.7$ & $0.78$\\ 
		%\hline
		Altitude (L) ($m$)   & $1797$ & $1791$ & $1768$\\ 
		%\hline
	    % Frequency  & high & high & high\\ 
		%\hline
	    Max. depth ($m$)   & $71$ & $77$ & $42$\\ 
		\hline
		Meteo station &  Segl Maria & Segl Maria & Samedan\\ 
		%\hline
		Dist. to lake ($km$)   & 0.5 & 1 & 5\\
		%\hline
	    Altitude (S) ($m$)  & $1804$ & $1804$ & $1709$\\ 
		\hline
	    \end{tabular}
	    \vspace{-0.5em}
	    \caption{Characteristics of the target
              lakes. Altitude (L) and altitude (S) denote the
              altitudes of the lake and nearest meteo station
              respectively. The distance to station is also shown.}
	    \label{table:lakes}
	\end{table}
\normalsize
\begin{table*}[th]
\centering
\small
\vspace{-0.5em}
\begin{tabular}{cccccccc}
    \hline
    \textbf{Lake} & \textbf{Winter} & \multicolumn{2}{c} {\textbf{Non-transition days}} & \textbf{Transition days} & \textbf{Total} & \textbf{Temporal resolution (days)}  & \textbf{\# lake pixels}\\
    % \hline
    % \textbf{Inactive Modes} & \textbf{Description}\\
    \cline{3-4}   &
    & \textbf{Non-frozen} & \textbf{Frozen} & & \textbf{\# acq.} & &     \\
    %\hhline{~--}
    \hline
    \multirow{2}{*}{Sils}       & 2016-17 & 40 & 42 & 37 & 119 & 2.3 & 40908 \\ %\hline
                                & 2017-18 & 76 & 65 & 40 & 181 & 1.5 & \\ \hline
    \multirow{2}{*}{Silvaplana} & 2016-17 & 36 & 44 & 39 & 119 & 2.3 & 26563 \\ %\hline
                                & 2017-18 & 85 & 66 & 30 & 181 & 1.5 & \\ \hline
    \multirow{2}{*}{St.~Moritz} & 2016-17 & 66 & 42 & 11 & 119 & 2.3 & 7521 \\ %\hline
                                & 2017-18 & 84 & 77 & 20 & 181 & 1.5 & \\ \hline
  \end{tabular}
  \vspace{-0.5em}
  \caption{Dataset statistics. Non-transition days, on which a
    lake is almost fully frozen / non-frozen, and transition days
    (partially frozen) are shown. Lake pixels are those which
    lie completely inside the lake polygon. \# acq. denotes the number of acquisitions.}
  \label{table:dataset_stats}
 \end{table*} 
\normalsize

\vspace{-1em}
\section{RELATED WORK}
%\subsection{Lake ice monitoring}
Many studies discussed the trends in lake ice formation in different
parts of the globe. \cite{Duguay2006journal} presented the trends in
lake freeze-up and break-up across Canada for a long period from
$1951$ until $2000$. Later, \citeauthor{franssen_scherrer2008} and Scherrer (2008)
reported the decreasing tendency in lake freezing in several Swiss
lakes. Then, \citeauthor{Brown_and_Duguay_2010} and Duguay (2010) reviewed the response and
role of ice cover in lake-climate interactions. This paper observed
that the ability to accurately monitor lake ice will be an important
step in the improvement of global circulation models, regional and
global climate models and numerical weather forecasting. \citeauthor{BrownDuguay_lakeice_2011} and Duguay (2011) used the Canadian Lake Ice Model (CLIMo) to simulate lake ice phenology across the North American Arctic from $1961$–$2100$, using two climate scenarios produced by the Canadian Regional Climate Model (CRCM). They projected changes to the ice cover using $30$-year mean data between $1961$–$1990$ and $2041$–$2070$, which suggested a probable drift in freeze-up ($0$–$15$ days later) and break-up ($10$–$25$ days earlier). \cite{Duguay_lake_river_ice2015} presented an overview of the progress of remote sensing for lake and river ice. For lakes, that work reviewed a number of topics, including ice cover concentration, ice extent and phenology, and ice types, as well as ice thickness, snow on ice, snow/ice surface temperature, and grounded and floating ice cover on shallow Arctic and sub-Arctic lakes.

%Recently, \cite{rs11161952} provided an  overview of recent achievements, challenges, and opportunities for land remote sensing of cold regions by (a) summarizing the physical principles and methods in remote sensing of selected key variables related to ice, snow, permafrost, water bodies, and vegetation; (b) highlighting recent environmental nonstationarity occurring in the Arctic, Tibetan Plateau, and Antarctica as detected from satellite observations; (c) discussing the limits of available remote sensing data and approaches for regional monitoring; and (d) exploring new opportunities from next-generation satellite missions and emerging methods for accurate, timely, and multi-scale mapping of cold regions

\textbf{Lake ice monitoring using Radar data.}  \citeauthor{Duguay_ice_depth_thickness_SAR_optical2003} and Lafleur (2003) proposed to
determine the depth and thickness of ice in shallow lakes and ponds
using the Landsat Thematic Mapper and European Remote Sensing (ERS)-1 SAR data. Almost a decade later, \cite{SurduDuguayBrown_2014} conducted a study of the shallow lakes on the north slope of Alaska to find the response of ice cover on the climate
conditions using Radar remote sensing and numerical analysis. A
machine learning-based automated ice-vs-water classification was
proposed by \cite{ML_SAR_classification_journal_2014} using dual
polarisation SAR imagery. Later, \cite{ice_SAR_Duguay_2015} performed
a study on the ice freezing and thawing detection in shallow lakes
from Northern Alaska with spaceborne SAR
imagery. \cite{Duguay_Kaab_2016} monitored ice phenology in lakes of
the Lena river delta using TerraSAR-X
backscatter. \cite{Duguay_satellite_NHemi_2017} performed an
assessment of lake ice phenology in the Northern Hemisphere from
$2002$ to $2015$. \cite{lakesize_lakeice_SAR2018} studied the effect
of the lake size on the accuracy of a threshold-based classification
of ground-fast and floating lake ice from Sentinel-1 SAR
data. %Recently, \cite{wetsnow_vs_drysnow_SAR_2018} put forward an interesting machine learning approach to detect wet and dry snow in mountainous areas using Sentinel-1 SAR data.
\citeauthor{Duguay2019_Sentinel1SAR} and Wang (2019) presented various algorithms
such as thresholding, Iterative Region Growing with Semantics (IRGS)
and $k$-means for the generation of a floating lake ice product from
Sentinel-1 SAR data for various permafrost regions (Alaska, Canada and
Russia).
%They found that the thresholding algorithm performed slightly
%better on average than the IRGS algorithm, which outperformed $k$-means.
\cite{Geldsetzer_2010_Journal} used RADARSAT-2 SAR data to monitor ice
cover in lakes during the spring melt period in the Yukon area of the
Canadian Arctic. They put forward a threshold-based classification
methodology and observed that the \textit{HH} and \textit{HV}
backscatter from the lake ice have significant temporal variability
and inter-lake diversity. \cite{RADARSAT2_Murfitt_Brown_2018} used the
RADARSAT-2 imagery to develop a threshold-based method to determine
lake phenology events for the mid-latitude lakes in Central Ontario
from $2008$ to $2017$. \cite{radarSAT2_Duguay_2018} also used
RADARSAT-2 imagery (dual polarised) for developing a lake ice
classification system acquired over lake Erie, with the IRGS
method. Additionally, \cite{RADARSAT2_Duguay_2018_journal} used the
polarimetric RADARSAT-2 (C-Band) to observe the scattering mechanisms
of bubbled freshwater lake ice. SAR data analysis is challenging, and deep learning could play a
significant role because of its ability to learn task-specific,
hierarchical image features. %\cite{seaice_SAR_Duguay_2017} used CNNs to estimate sea ice concentration using SAR data acquired during freeze-up period in the Gulf of St.\ Lawrence on the east coast of Canada.
%To our knowledge, deep learning has not yet been used for lake ice monitoring with Radar data.

\textbf{Lake ice monitoring with webcams.}  \cite{muyan_lakeice_2018} described a system
that detects lake ice in webcam data with the help of a deep neural
network.
Public webcams have two main advantages compared to optical satellite
images. Firstly, they are usually not affected by clouds, except for
the comparatively rare case of dense fog. Secondly, they have a very
high temporal resolution (up to one image per 10 min). Although the
approach generated excellent results, it also has
disadvantages. Webcams are usually placed arbitrarily (e.g., too far
away or covering a small lake area), and often only low above the
lake, leading to great scale differences between front and back of the
lake surface. Moreover, they are prone to hardware failure, and, being
very cheap cameras, they have poor spectral and radiometric quality
with significant compression artifacts. Another practical problem with
webcams is that it is difficult to operationalise them at
country- or even world-scale.
%In addition, multiple webcams are needed
%to get full coverage of a lake, especially for big lakes.

\textbf{Optical data for lake ice monitoring.}  \cite{tom_lakeice_2018} proposed a
machine learning-based methodology for lake ice detection using low
resolution optical satellite images. The main problem with optical
satellite images is the data loss due to clouds. However, the authors
showed that the algorithm produces consistent results when tested on
data from multiple winters. In addition, \cite{Barbieux_2018_Journal}
used Landsat-8 multi-spectral data for extraction of frozen lakes and
water-vs-ice classification. Recently, \cite{LIP1_final_report_2019}
put forward a feasibility study, which targeted for a unified lake ice
monitoring system that combines images from optical satellites,
in-situ temperature data and webcam images.
%\par
%\subsection{Radar data and deep learning} Later, \cite{DL_SAR_scene_classification_journal_2019} proposed a Feature Re-calibration Network with Multi-scale Spatial Features (FRN-MSF), to achieve high accuracy in SAR-based scene classification. They achieved promising results on different types of SAR scenes. In addition, \cite{Krestenitis_2019} identified oil spills using SAR data and deep learning and performed semantic segmentation with deep CNNs where \textit{deeplabv3+} \citep{deeplabv3plus2018} recorded the best performance, in terms of test set accuracy and inference time.
%\vspace{-1em}

\vspace{-0.25em}
\section{data}\label{sec:data}
\begin{figure}
    \begin{center}
	    \includegraphics[width=0.75\columnwidth]{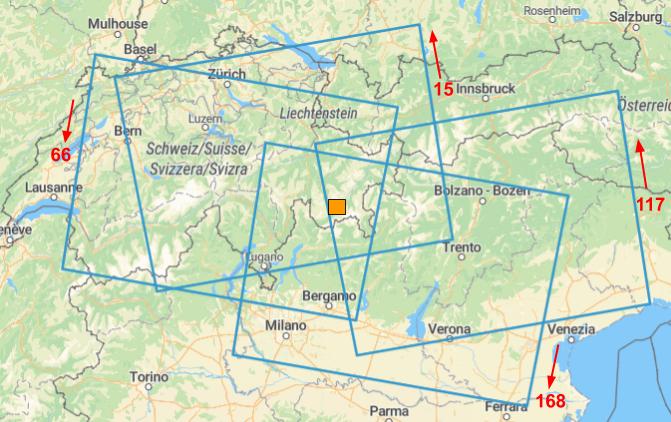}
	    \vspace{-1em}
	\caption{The four Sentinel-1 orbits ($15$, $66$, $17$, $168$)
          with the corresponding directions (\textit{ascending} or
          \textit{descending}) which covers the \textit{Region Sils}
          (shown as orange filled rectangle).}
    \label{fig:4orbits}
    \end{center}
\end{figure}

\textbf{Sentinel-1 SAR} consists of two identical satellites
(\textit{S1A} and \textit{S1B}) operational in space with $180^\circ$
phase shift, following a sun-synchronous, near-polar orbit. The two
satellites orbit the Earth at an altitude of $693$ km and have a
repeat cycle of $12$ days at the equator (effectively $6$ days with
\textit{S1A} and \textit{S1B}). The same point on Earth is mapped
several times within one repeat cycle. Due to the large across-track
area coverage of the satellites and the latitude of our target area in
Switzerland (and most other areas where lakes freeze), the revisit
time is further reduced. For Region Sils, it can bee seen from Table
\ref{table:dataset_stats} that the temporal resolution in winter
$2017$-$18$ is better than that of $2016$-$17$. This is because of missing
data from \textit{S1B}. Though \textit{S1B} was launched in April
2016, the corresponding data is fully available in the Google Earth Engine (GEE) platform
(see Section ``Data Collection'' below) only from March $2017$. In addition, Sentinel-1 covers the Region Sils in four orbits. See Table \ref{tab:orbits} for the details. Footprints of the four orbits are shown in Fig. \ref{fig:4orbits}.
\begin{table}
\centering
\small
\caption{Details of four orbits scanning \textit{Region Sils} such as orbit number, flight direction, acquisition time in Universal Coordinate Time (UTC), and approximate incidence angle.}%
\vspace{-0.5em}
\begin{tabular}{|cccc|}
\hline
\textbf{Orbit} & \textbf{Flight dir.} & \textbf{Acquisition time} & \textbf{Incidence angle} \\ \hline
15 & ascending & 17:15 & 41.0$^{\circ}$ \\
%\hline
66 & descending & 05:35 &  32.3$^{\circ}$ \\
% \hline
117 & ascending & 17:06  &  30.8$^{\circ}$\\
168 & descending & 05:26 & 41.7$^{\circ}$ \\
\hline
\end{tabular}
\label{tab:orbits} 
\end{table}
\normalsize

\begin{figure}[!ht]
    \centering
    \subfigure[Distribution of VV]{\includegraphics[width=0.45\textwidth]{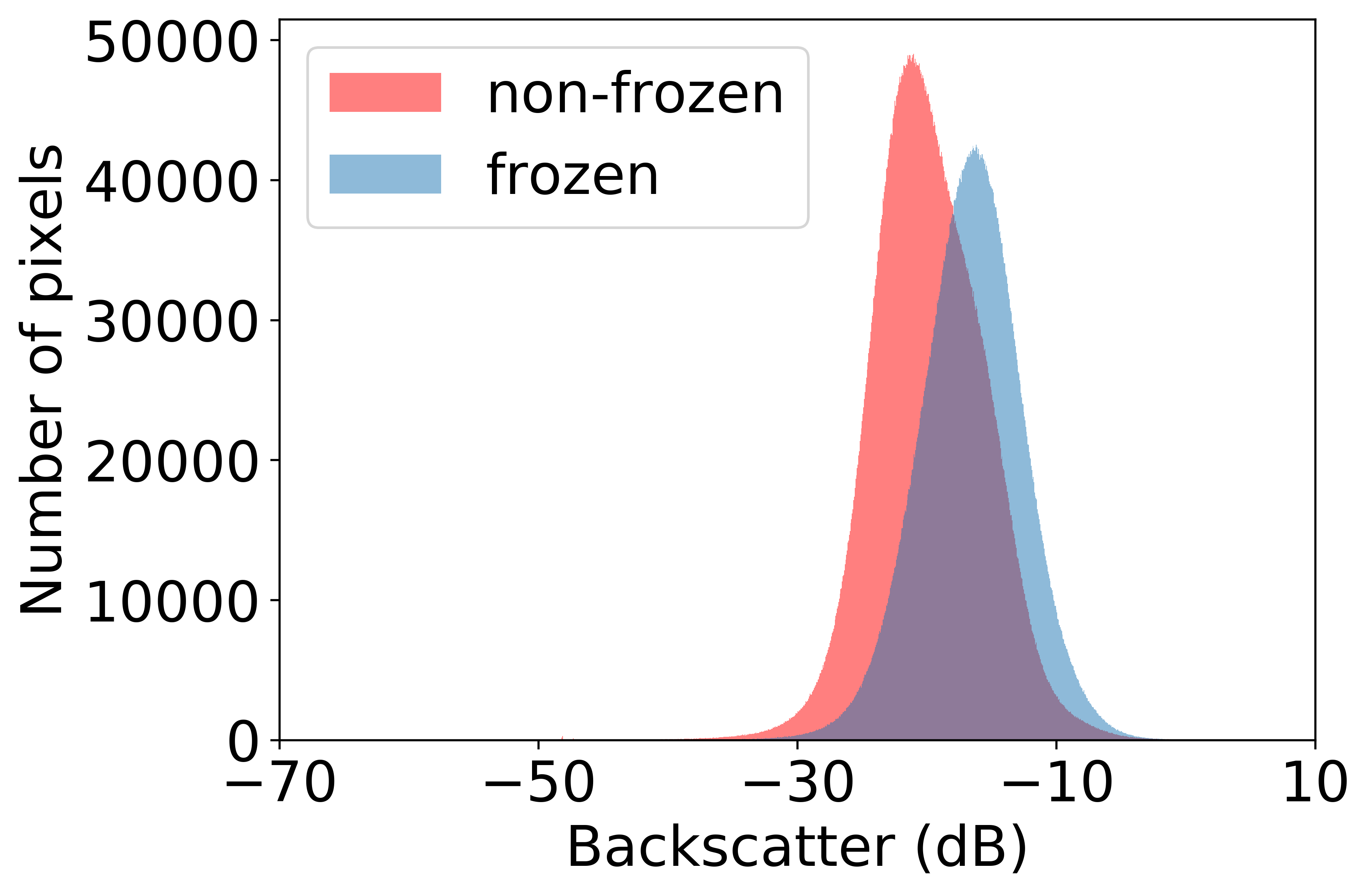}}
    \subfigure[Distribution of VH]{\includegraphics[width=0.45\textwidth]{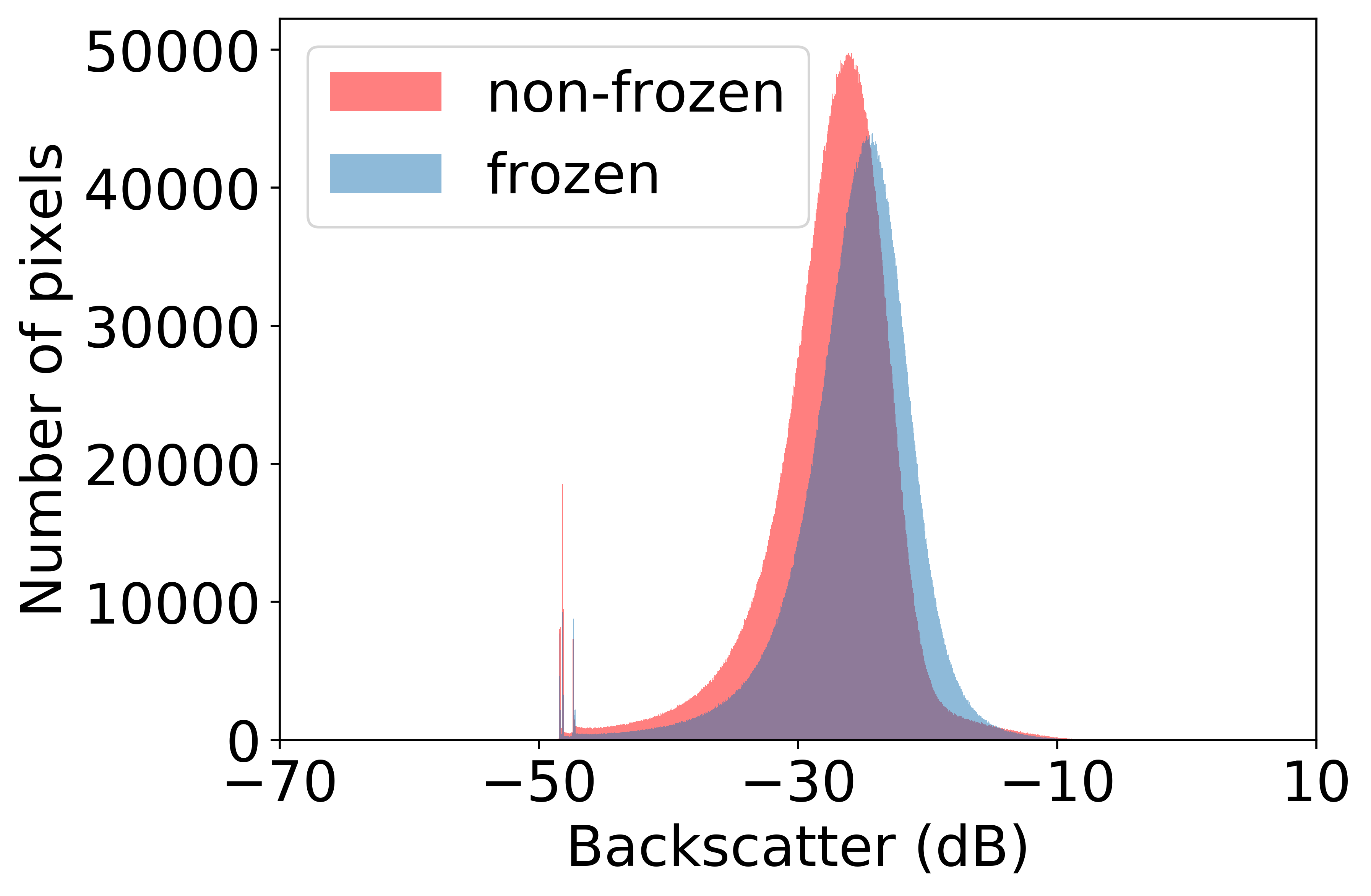}}
    \vspace{-1em}
    \caption{Distribution of frozen and non-frozen pixels for VV and
      VH polarisations (combined data from $3$ lakes, $2$ winters). Best if viewed on screen.}
\label{fig:distribution}
\end{figure}

The \textit{S1A} and \textit{S1B} satellites have the same SAR system
on board which sends out frequency-modulated electromagnetic waves (C-frequency band) and detects the backscattered echoes of the surface. From the reflected signal, the SAR sensor measures the
amplitude and phase. In our research, we use only the amplitude
information. We work with the Level-1 Ground Range Detected (GRD)
product in Interferometric Wide (IW) swath mode. That product has no
phase information anymore, and has a nearly square footprint
($10$m$\times10$m per pixel). It also has good temporal resolution
(see Table \ref{table:dataset_stats}) and four polarisations
(\textit{VV, VH, HH, HV}). However, for the Region Sils,
data is acquired only in \textit{VV} and \textit{VH} modes. The distribution of backscatter
values of \textit{frozen} and \textit{non-frozen} pixels in these
bands are shown in Fig.~\ref{fig:distribution}. Note that the
separability in \textit{VV} appears better than in \textit{VH}.

The radar backscatter is influenced in a complex manner by a variety of factors, which can be
grouped into two main categories. Firstly, the \textit{sensor
  parameters} such as wavelength ($5.54$ cm), incident angle
($20^{\circ}$ to $46^{\circ}$) and polarisation. Secondly, it depends
on \textit{surface parameters} which can be either geometrical factors
such as roughness, landscape topography, etc.\ or physical factors such
as the permittivity of the surface material. Significant
factors for lakes are also wind speed and direction, and the water content in snow. For smooth and plain water, almost all radiation is scattered away from the sensor making it appear very dark. As the wind speed picks up, waves occur on the water surface and significant scattering can occur. When perfectly plain water is covered by perfectly plain ice, microwaves penetrate the ice without absorption and are reflected at the ice-water interface away from the sensor, and the ice covered lake would in theory appear completely black. In reality, cracks in the ice scatter some microwaves back to the sensor. Therefore, visible and well located cracks are clear indicators for ice cover. Furthermore, in reality, the ice-water interface is never completely smooth, therefore some scattering can occur at these boundaries which, however, can be weak. The older the ice, the more air bubbles are enclosed in it, which increase the backscatter within the ice volume by direct backscattering and also by double reflection of microwaves at the air-bubbles and the ice-water interface. With snow cover, the air-ice interface becomes increasingly rough which further increases the backscatter signal. Finally, with snow melt, the liquid water content of the overlying snow pack increases and significantly reduces the backscatter signal, as the snow-water mixture of wet snow absorbs a significant fraction of the microwave energy.
\begin{figure}
	\centering
	\subfigure[VV, wind speed <5 Km/h]{\includegraphics[width=0.45\textwidth]{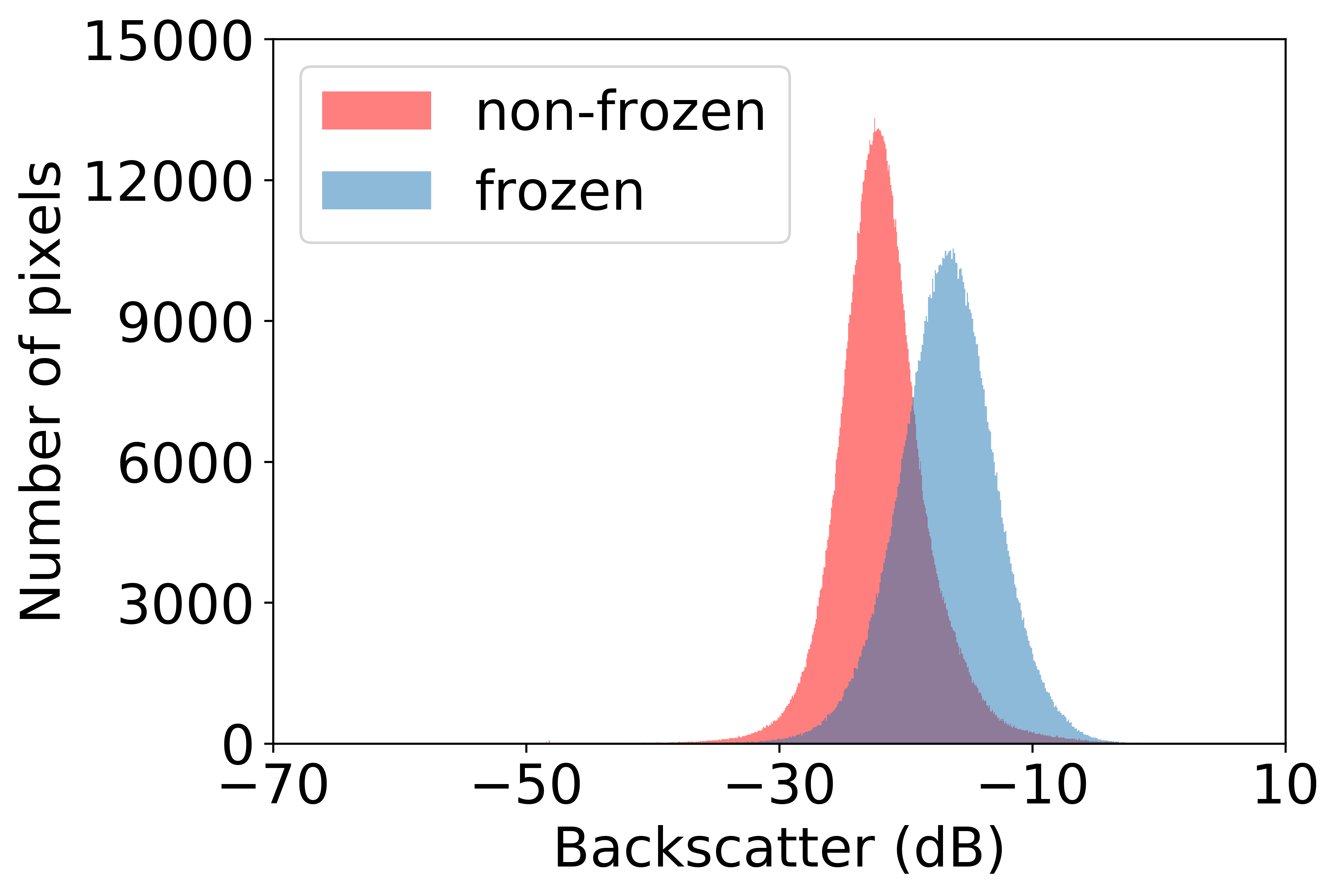}}
	\subfigure[VH, wind speed <5 Km/h]{\includegraphics[width=0.45\textwidth]{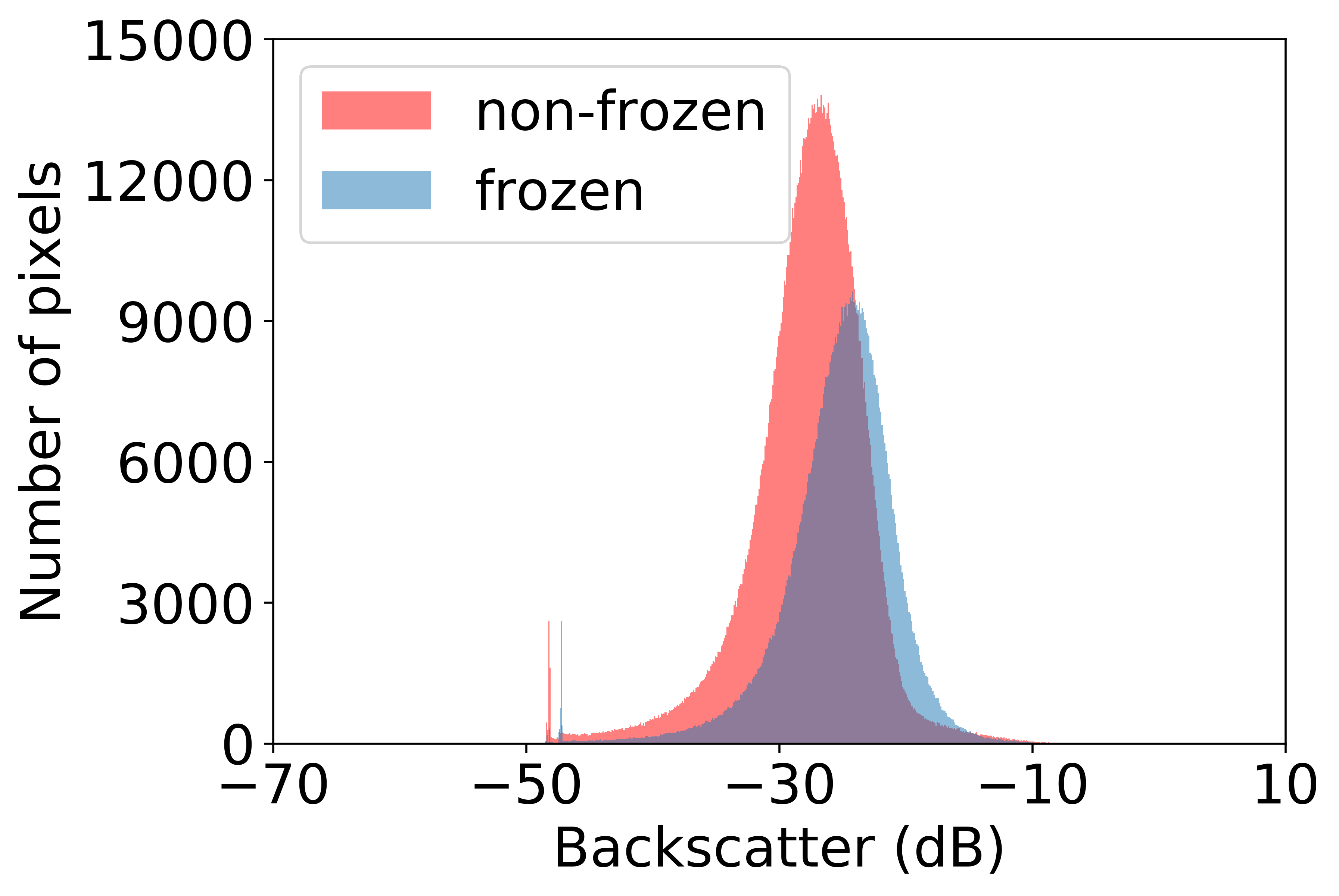}}
	\subfigure[VV, wind speed >20 Km/h]{\includegraphics[width=0.45\textwidth]{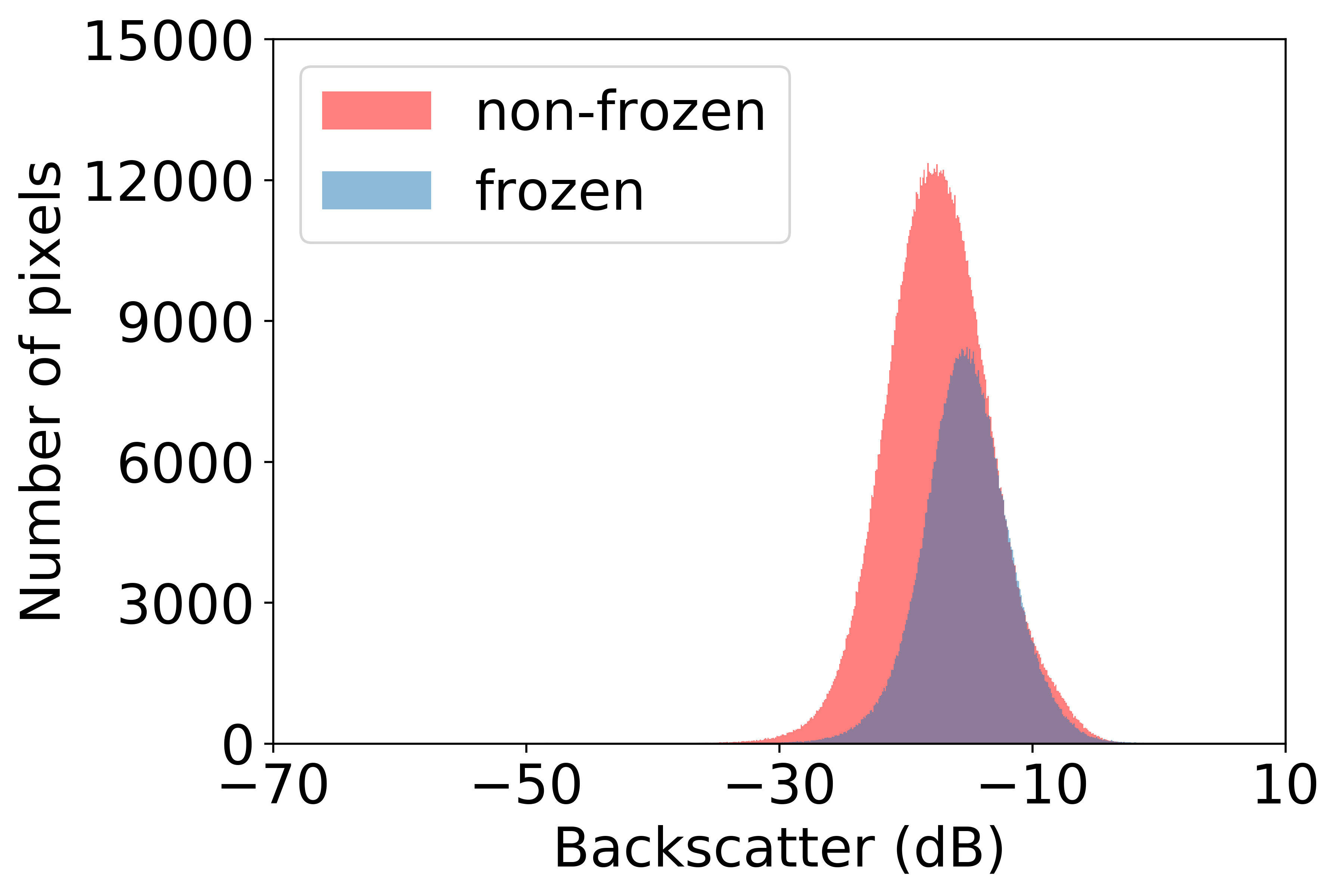}}
	\subfigure[VH, wind speed >20 Km/h]{\includegraphics[width=0.45\textwidth]{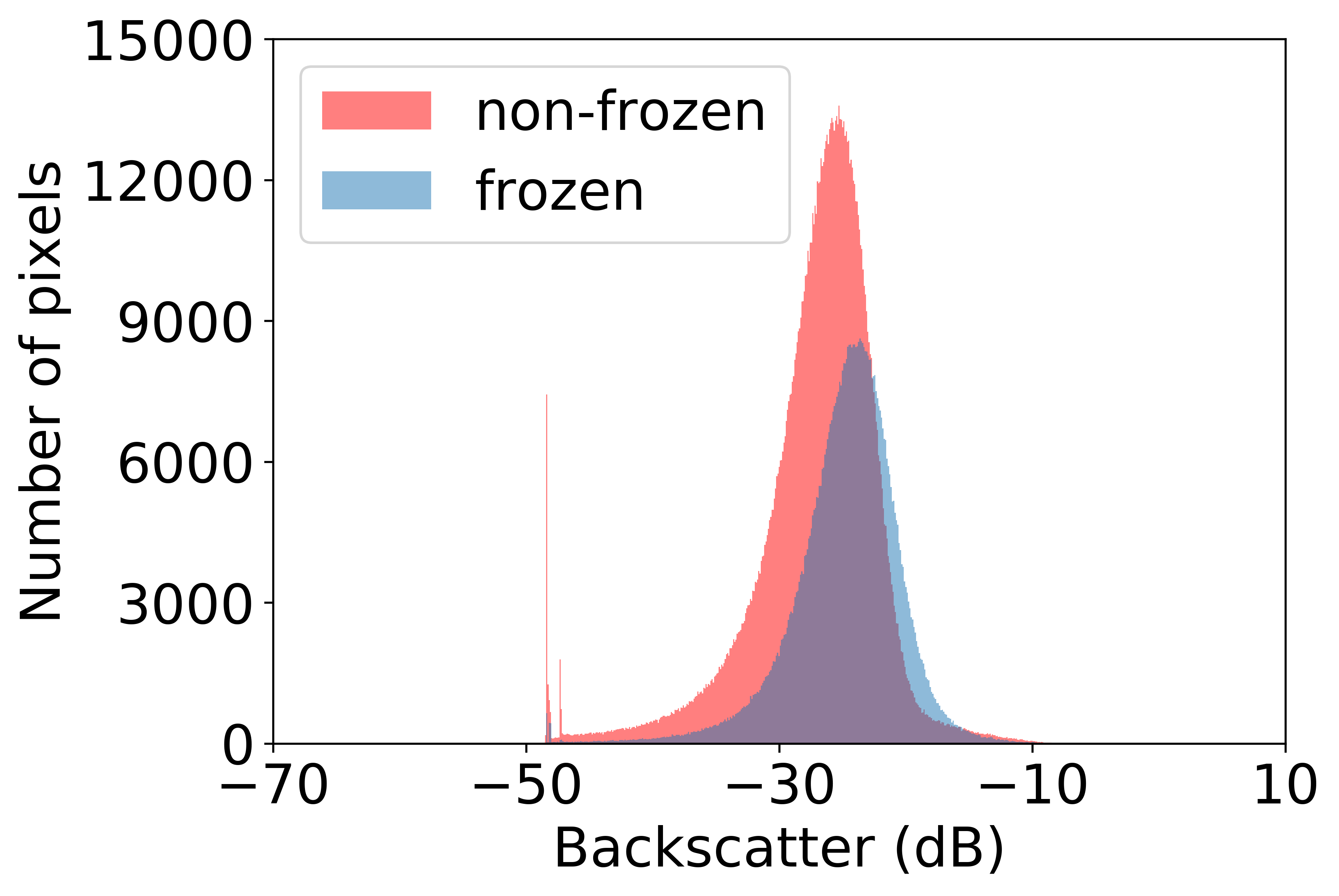}}
	\vspace{-1em}
	\caption{Distribution of frozen and non-frozen pixels for VV and VH polarisations in different wind speed conditions (combined data from $3$ lakes, $2$ winters). Best if viewed on screen.}	
	\label{fig:dist_wind}
\end{figure}
\par
\textbf{Data collection and pre-processing.}\label{sec:DataCollect_GEE}
GEE is a cloud-based platform for
large-scale geo-spatial data analysis \citep{gorelick2017google}.
%Large scale
%computing is ubiquitously possible in GEE which makes the
%planetary-scale remote sensing analysis uncomplicated, especially the
%pre-processing steps.
It stores and provides data of various satellite missions, performs
data pre-processing and makes them freely available for education and
research purposes. The Sentinel-1 backscatter coefficients (in
decibels) were downloaded from the GEE platform after several inbuilt
pre-processing steps such as GRD \textit{border noise removal} which
corrects the noise at the border of the images, \textit{thermal noise removal} for correcting the thermal noise between the sub-swaths, \textit{radiometric calibration} which calculates the backscatter intensity using the GRD metadata, \textit{terrain correction} to
correct the side looking effects using the digital elevation model
(SRTM, 30m), and \textit{log-scaling} to transform the approximate
distribution of the SAR responses from \textit{Chi-squared} to
\textit{Gaussian} (see Fig. \ref{fig:distribution}). Note that we
did not perform any absolute geolocation correction, since the
back-projected lake outlines suggested a sufficient accuracy.

\textbf{Transition and non-transition days.}  All the data from two
winters was divided into two categories: \textit{non-transition} dates
where the lake is almost fully frozen or fully non-frozen, and \textit{transition} dates with partially frozen lake surface. Both freeze-up and break-up dates belong to the transition category. The dataset statistics are shown in Table \ref{table:dataset_stats}. Note that, since lake St.~Moritz is relatively small in area and volume, it freezes and melts faster and has fewer transition days.

\textbf{Ground truth.} For each lake, one label (\textit{fully frozen / non-frozen}, \textit{partially frozen / non-frozen}) per day was assigned by a human operator after visual
interpretation of the visible part of the lake from freely available webcam data.  The ground truth thus generated was further enriched by visual
interpretation of the Sentinel-2 images whenever available. However,
some remaining noise in the ground truth is likely due to
interpretation errors, as a result of overly oblique viewing angles of
webcams and compression artefacts in the images. 
During the transition days, ground truth estimation was very difficult since we had partially frozen and non-frozen states and there was a difficulty to discriminate transparent ice and water. Thus, transition days were not used for quantitative analysis. %Note also, we have the ground truth available only for the non-transition days.
%Note also, the
%interpretation of complicated cases %(e.g. partially frozen dates) does
%exhibit certain level of discrepancy 5between different human
%operators.
%
%\textcolor{red}{UNCLEAR: what happenend with the transition days? You
%  cannot assign a single label there...}
%
\vspace{-1em}

\section{Methodology}\label{sec:METHODOLOGY}
\textbf{Semantic segmentation.}\label{sec:Semantic} We define lake ice
detection as a two class (\textit{frozen, non-frozen}) pixel-wise
classification problem and tackle it with the state-of-the-art
semantic segmentation network \textit{Deeplab v3+}
\citep{deeplabv3plus2018}. The \textit{non-frozen} class comprises of
only \textit{water} pixels. Whereas a pixel is considered to be part
of the \textit{frozen} class if it is either \textit{ice} or
\textit{snow}, since in the target region, the frozen lakes are covered by
snow for much of the winter. The standard procedures in machine
learning-based data analysis are followed. The dataset is first
divided into mutually exclusive \textit{training} and \textit{test} sets. The \textit{Deeplab v3+} model is then fitted on the training set. %This model is further used to predict the images in the validation set, which provides an unbiased evaluation of the model's predictive performance outside the training set while tuning the hyper-parameters.
Lastly, the trained model is tested on the previously unseen test dataset.

\begin{figure}[!ht]
	    \includegraphics[width=1.0\columnwidth]{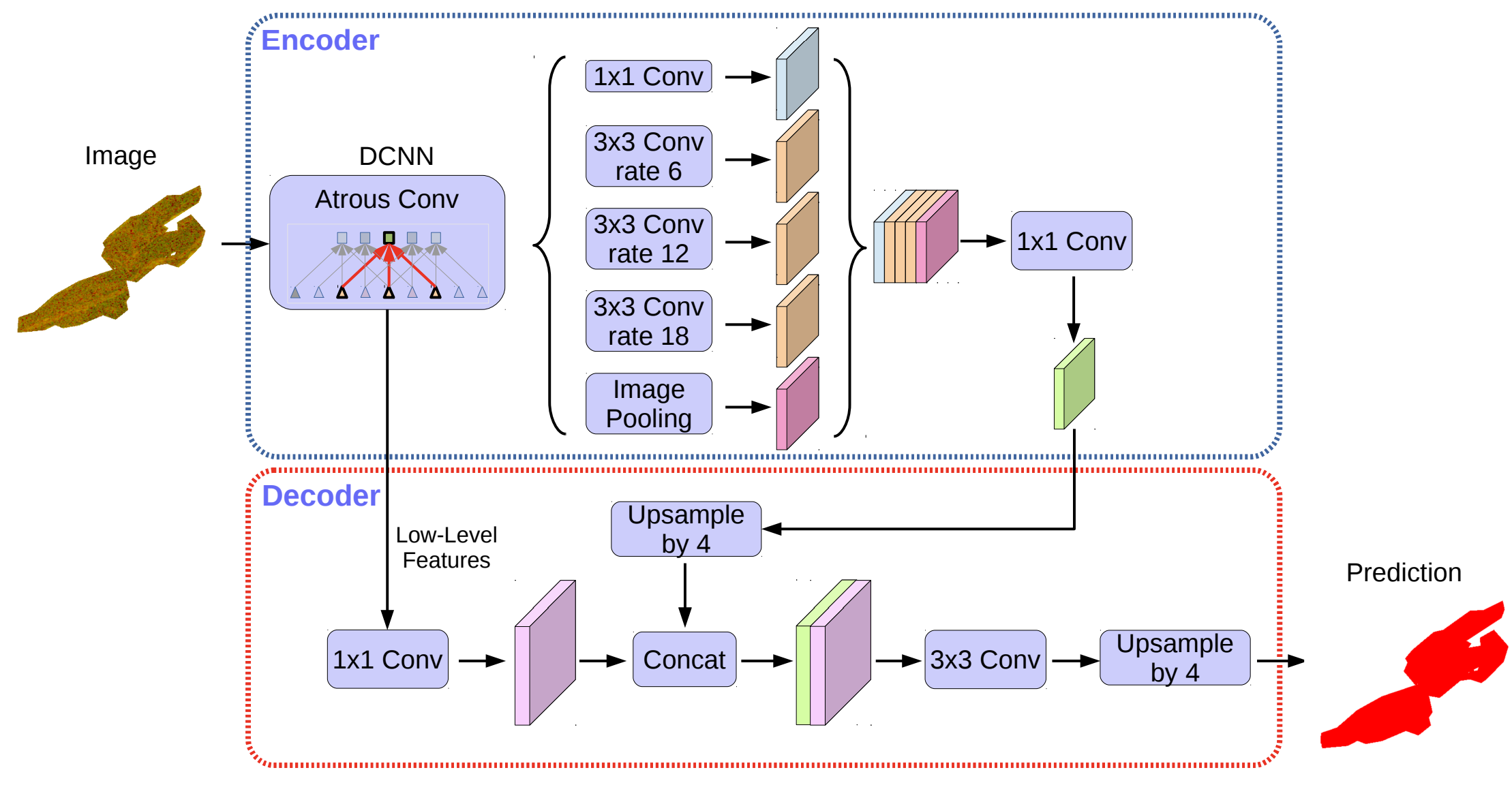}
	    \vspace{-1.5em}
	\caption{\textit{Deeplab v3+} architecture. Best if viewed on screen.}
    \label{fig:deeplabarch}
\end{figure}
\begin{table*}
    \centering
    \small
    \vspace{-0.5em}
    \begin{minipage}{.49\linewidth}
      %\caption{Leave W2016-17 out}
      \centering
        \begin{tabular}{|l|c|c|c|}
            \hline
            \diagbox[width=8em]{True}{Prediction} & Non-frozen~~~Frozen  & \textbf{Recall} \\
            \hline
            Non-frozen & 3.06~~~~~~~~~\textcolor{gray}{0.01} & 99.7\% \\ %\hline
            Frozen & \textcolor{gray}{0.29}~~~~~~~~~2.89 & 90.8\% \\ \hline
            \textbf{Precision} & 91.3\%~~~~~~99.6\%  & \textbf{95.5\%}\\
            \hline
        \end{tabular}
    \end{minipage}%
    \begin{minipage}{.49\linewidth}
      \centering
       % \caption{Leave W2017-18 out.}
        \begin{tabular}{|l|c|c|c|}\hline
            \diagbox[width=8em]{True}{Prediction} & Non-frozen~~~Frozen  & \textbf{Recall} \\ 
            \hline
            Non-frozen & 5.77~~~~~~~~~\textcolor{gray}{0.19} & 96.7\% \\ %\hline
            Frozen & \textcolor{gray}{0.44}~~~~~~~~~4.59 & 91.1\% \\
            \hline
            \textbf{Precision} & 92.9\%~~~~~~95.8\%  & \textbf{94.8\%}\\ \hline
        \end{tabular}
    \end{minipage} 
    \vspace{-0.5em}
    \caption{Results for winter $2016$-$17$ (left) and $2017$-$18$
      (right). Data from all the three lakes from winter $2016$-$17$ was
      used to train the model that was tested on winter $2017$-$18$, and
      vice versa. Confusion matrices are shown. Units are in millions
      of pixels, except for precision, recall, and \textbf{accuracy}
      (bottom right cell in each table).}
    \label{table:LOOCV_winter}
\vspace{0.5em}
\end{table*}
\normalsize

\textbf{Deeplab v3+} \citep{deeplabv3plus2018} is a deep neural
network for semantic segmentation, which has set the state-of-the-art
in multiple benchmarks, including among others
the PASCAL VOC 2012 dataset \citep{Everingham15}. It combines the advantages of both Atrous
Spatial Pyramid Pooling (ASPP) and encoder-decoder structure.
Atrous convolution allows to explicitly control the resolution of
the features computed by the convolutional feature
extractor. Moreover, it adjusts the field-of-view of the filters in
order to capture multi-scale information. \textit{Deeplab v3+} also
incorporates depthwise separable convolution (per-channel 2D
convolution followed by pointwise $1\times1$ convolution) which
significantly reduces the model size. The network architecture is shown in Fig.~\ref{fig:deeplabarch}.

\textbf{Network parameters.}  We used the \textit{mobilenetv2}
implementation of \textit{Deeplab v3+}, as available in
\emph{TensorFlow}. The \textit{train crop size} was set to
$129\times129$ (effective patch size is $128\times128$) and the
\textit{eval crop size} to the full image resolution. All models were
trained for $40'000$ iterations with a batch size of $8$. Atrous rates
were set to [$1, 2, 3$] for all experiments. The cross-entropy loss
function was minimised with standard stochastic gradient descent, with
a base learning rate of $10^{-3}$.

\textbf{Transfer learning.}  Deep supervised classification approaches
need lots of labelled data and a large amount of resources to
train a model from scratch. Such data volumes are often not
available. Even if they are, labelling them is costly and increases
the computational cost of model training.  \textit{Transfer learning}
mitigates this bottleneck by using an already trained model from some
related task as a starting point.  Given the fact that the initial
layers of a neural network learn rather generic local image
properties, a model trained on a huge image dataset can be re-utilised
on a different dataset with a much smaller amount of fine-tuning
(re-training) to the specific characteristics of the new data. We use
a \textit{Deeplab v3+} model pre-trained on the PASCAL VOC 2012
close-range dataset as the starting point and fine-tune it on the
relatively small Sentinel-1 SAR dataset (see Table
\ref{table:dataset_stats}). Surprisingly, we find that pre-training
on RGB amateur images of indoor scenes, vehicles, animals, humans etc.\ greatly enhances the performance even on a data source as different as interferometric Radar, compared to training from scratch only on the SAR data. Note, all weights were fine-tuned, we did not freeze any network layers.
%\subsection{Multi-Temporal Analysis (MTA)}
%\vspace{-0.75em}

\section{EXPERIMENTS, RESULTS, AND DISCUSSION}
\label{sec:EXPERIMENTS}
%\subsection{Evaluation metrics}
%We report recall, precision, overall accuracy and the Intersection Over Union (IOU) values.
%
We use various measures to quantify performance, including recall,
precision, overall accuracy, and the IoU score (Jaccard index). In all
experiments described in the paper, we used only \textit{lake pixels}
from the non-transition dates to train the network and to compute
performance metrics. This is due to the lack of reliable ground
truth during the transition dates. Additionally, whenever the ground truth 
cannot be established for a non-transition date due to foggy webcam images and/or clouds 
in Sentinel-2, it is exempted from the training set. 
%Note that, this strategy
%ensures the presence of only reliable reference data in the training
%set.
%Note also, the quantitative analysis is done only on the
%non-transition dates.
However, qualitative analysis is done on all the dates.
\par
\textbf{Quantitative results: Semantic segmentation.}  For
developing an ideal operational system for lake ice monitoring, the
data from a couple of lakes from a few winters would have to be used to train
the model, which can then be deployed on unseen lakes and
winters. However, generating ground truth for each lake is a tedious
task. Nevertheless, we make sure that the data from at least one lake
from one full winter is in the training set for the classifier to
learn the proper class decision boundaries.

%\textbf{Cross validation (CV).}\label{sec:LOOCV}
We employ Cross-Validation (CV), i.e., the data is partitioned into
\textit{k} folds, usually of approximately the same size. Then, the
evaluation is done \textit{k} times, each time using one fold as test
set and the union of all remaining folds as training set.
\textit{Leave-one-out cross-validation} is the setting where the
number of folds equals the number of instances (in our case the number
of winters/lakes) in the dataset.
\begin{figure*}
\centering
    \subfigure{\includegraphics[width=0.27\textwidth]{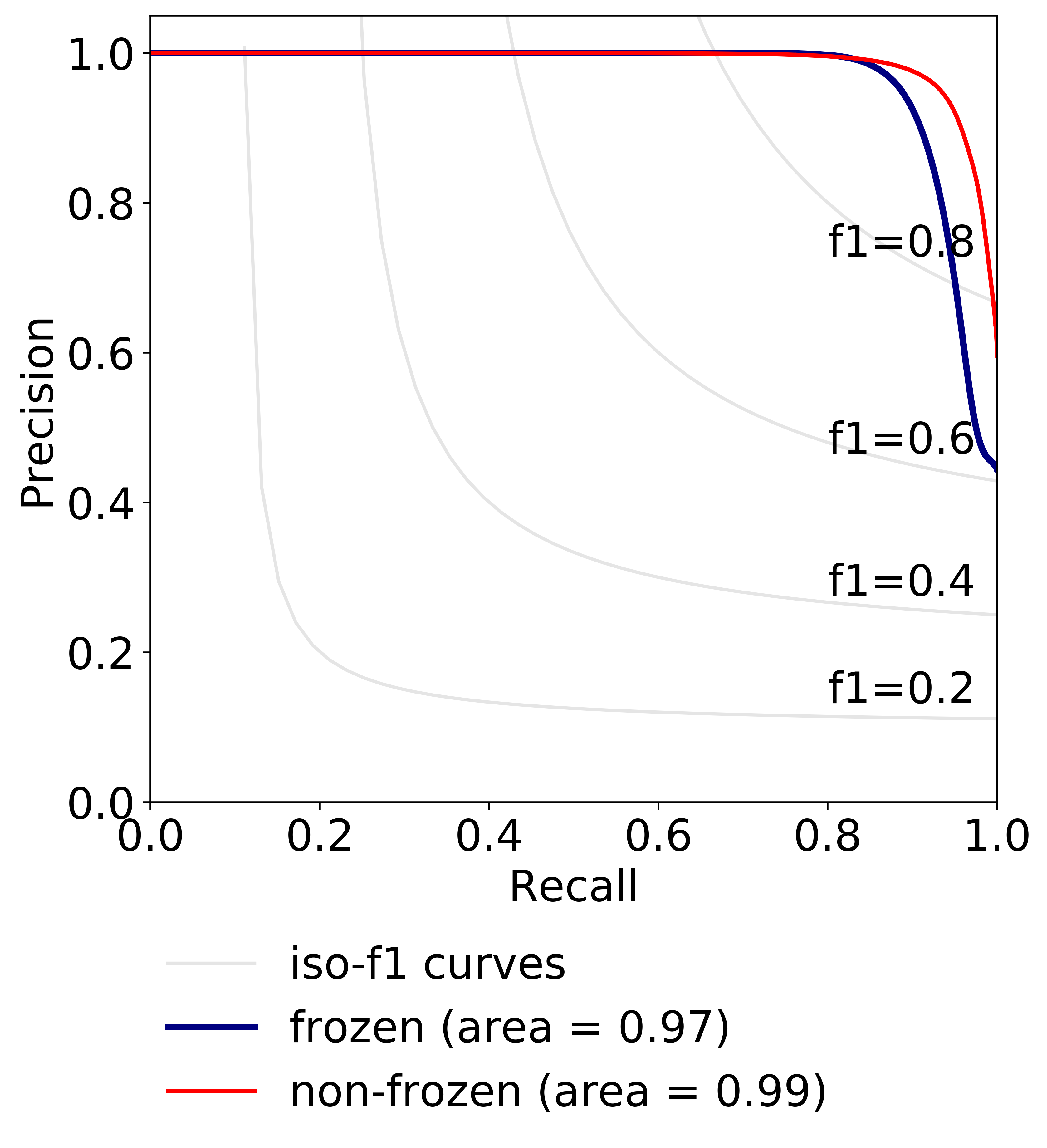}}
    \subfigure{\includegraphics[width=0.27\textwidth]{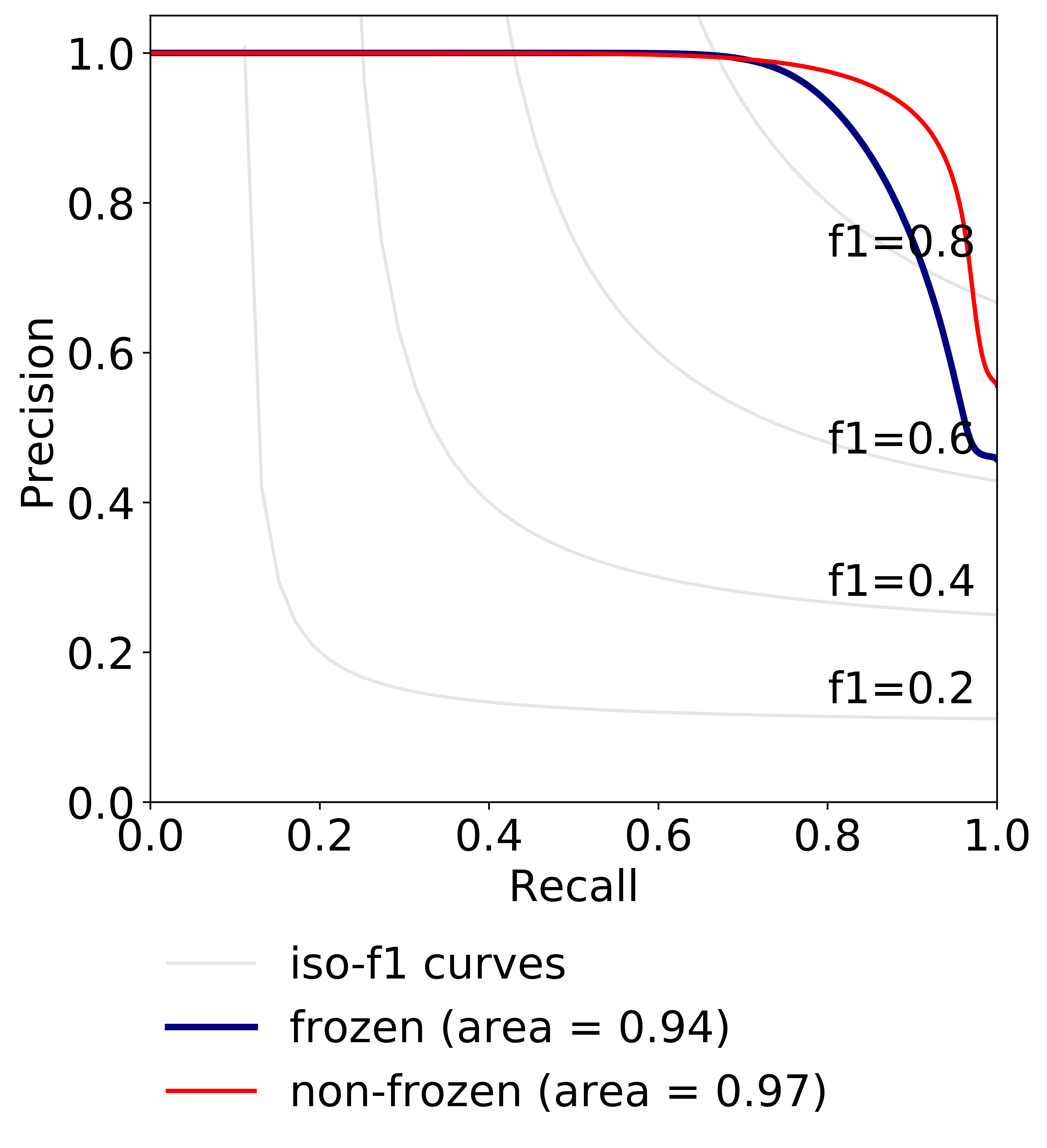}}
    \subfigure{\includegraphics[width=0.27\textwidth]{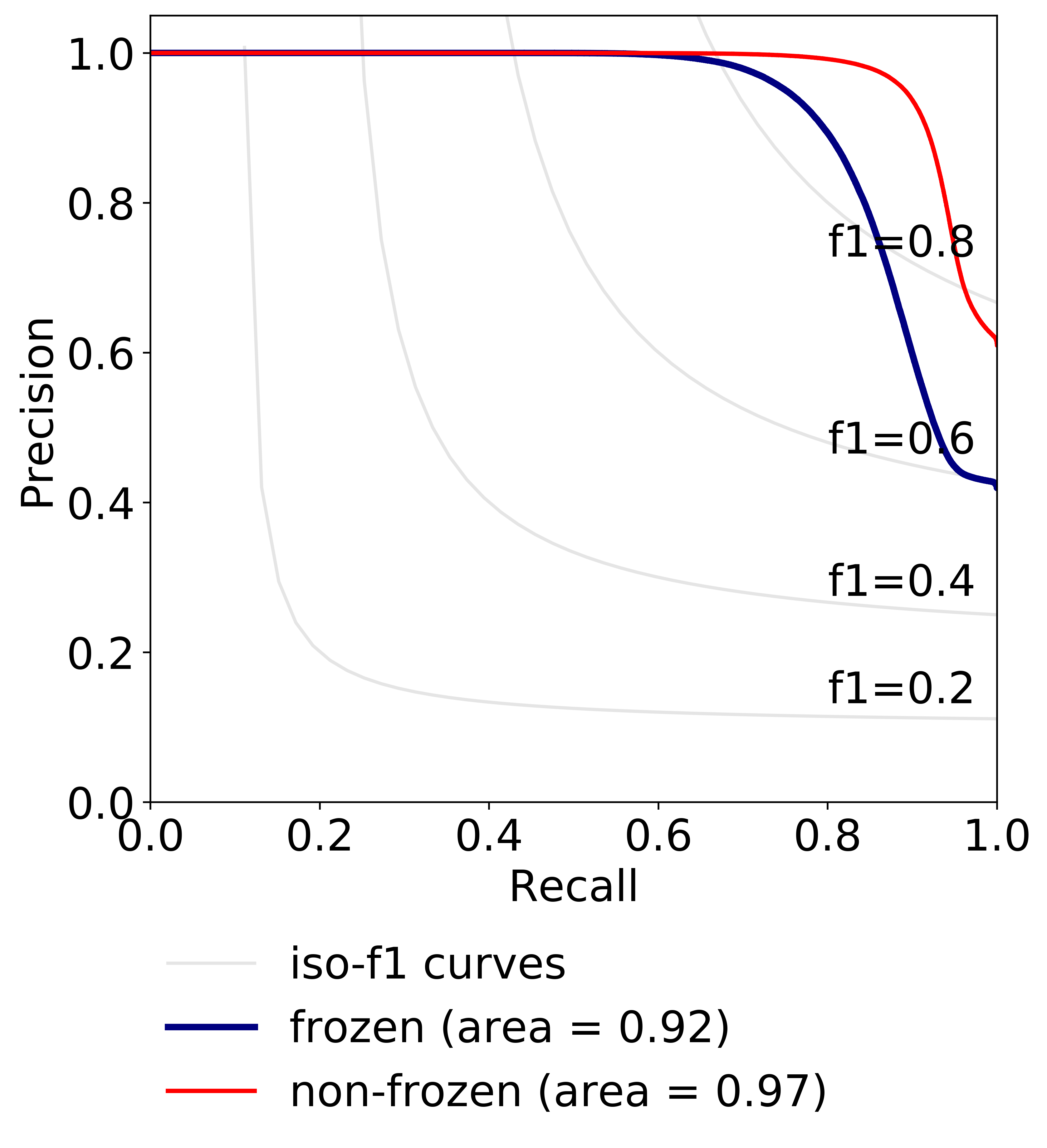}}
    \vspace{-1.5em}
    \caption{Precision-recall (PR) curves for lakes Sils (left), Silvaplana
      (middle), and St.~Moritz (right) for winter $2017$-$18$. The \textit{iso-f$1$} curve
      connects all points in the PR space with same
      \textit{F$1$} score. Combined data of all $3$ lakes from winter $2016$-$17$ was used to train the model. Best if viewed on screen.}
\label{fig:PR_curves}
\end{figure*}
The goal of \textit{leave-one-winter-out CV} is to investigate the
generalisation capability of a model trained on one winter when tested
on a different winter. The results are shown on Table
\ref{table:LOOCV_winter}. It can be seen that we achieve excellent
results for both winters with average accuracies of $95.5\%$ and
$94.8\%$ for $2016$-$17$ and $2017$-$18$ respectively. The
results show that the model generalises well across the
potential domain shift caused by the specific conditions of different
winters, without having seen data from any day within the test period.
\begin{table}[!ht]
    \centering
    \small
    \vspace{-0.5em}
    \begin{minipage}{\linewidth}
      %\caption{Leave Sils out}
      \centering
        \begin{tabular}{|l|c|c|c|}\hline
            \diagbox[width=8em]{True}{Prediction} & Non-frozen~~~Frozen  & \textbf{Recall} \\
            \hline
            Non-frozen & 4.69~~~~~~~~~\textcolor{gray}{0.02} & 99.4\% \\ %\hline
            Frozen & \textcolor{gray}{0.13}~~~~~~~~~4.22 & 96.9\% \\
            \hline
            \textbf{Precision} & 97.3\%~~~~~~99.4\%  & \textbf{98.3\%}\\ \hline
        \end{tabular}
    \end{minipage}\\
    \vspace{0.5mm}
    \begin{minipage}{\linewidth}
      \centering
        %\caption{Leave Silvaplana out.}
        \begin{tabular}{|l|c|c|c|}\hline
            \diagbox[width=8em]{True}{Prediction} & Non-frozen~~~Frozen  & \textbf{Recall} \\
            \hline
            Non-frozen & 3.08~~~~~~~~~\textcolor{gray}{0.10} & 96.4\% \\ %\hline
            Frozen & \textcolor{gray}{0.11}~~~~~~~~~2.78 & 96.1\% \\
            \hline
            \textbf{Precision} & 96.4\%~~~~~~96.4\%  & \textbf{96.4\% }\\ \hline
        \end{tabular}
    \end{minipage}\\
    \vspace{0.5mm}
    \begin{minipage}{\linewidth}
      \centering
        %\caption{Leave St.~Moritz out.}
        \begin{tabular}{|l|c|c|c|}\hline
            \diagbox[width=8em]{True}{Prediction} & Non-frozen~~~Frozen  & \textbf{Recall} \\
            \hline
            Non-frozen & 1.00~~~~~~~~~\textcolor{gray}{0.10} & 94.1\% \\ %\hline
            Frozen & \textcolor{gray}{0.62}~~~~~~~~~0.82 & 88.5\% \\
            \hline
            \textbf{Precision} & 94.1\%~~~~~~88.5\%  & \textbf{91.5\% } \\ \hline
        \end{tabular}
    \end{minipage}
    \vspace{-0.5em}
    \caption{Results for lake Sils (top), Silvaplana (middle), and
      St.~Moritz (bottom). Confusion matrices are shown for the
      leave-one-lake-out cross-validation experiment. Units are in
      millions of pixels, except for precision, recall, and
      \textbf{accuracy} (bottom right cell in each table).}
    \label{table:LOOCV_lakes}
\end{table}
\normalsize

We also report results of a \textit{leave-one-lake-out} CV experiment to check the generalisation capacity of the model across lakes. The results are shown on Table \ref{table:LOOCV_lakes}. While testing for all data of a lake (e.g., \textit{Sils}) from two winters, the data from the other two lakes (e.g., \textit{Silvaplana, St.~Moritz}) from the same two winters is used for training. The prediction achieves \textgreater$91.5\%$ overall accuracy for all three lakes. See Table
\ref{tab:iou_lakes} for the per-class and mean IoU values for each lake. The worst performance is noted for St.~Moritz where the lake itself is smaller than the size of a single patch ($128\times128$). Note also, St.~Moritz has the least mIoU, especially for class \emph{frozen}. This is because of the presence of tents and other infrastructure on the frozen lake St.~Moritz, which is not present in the other two lakes used in training. To assess the per-class performance in detail, we also report the precision-recall curves in Fig.~\ref{fig:PR_curves}. For both the frozen and non-frozen classes, the area under the curve is nearly optimal for lake Sils and very good performance is achieved on lakes Silvaplana and St.~Moritz.
\begin{table}[]
\centering
\small
\caption{Per-class- and mean IoU values of frozen and non-frozen classes for each lake. The data of a lake from two winters ($2016-17$ and $2017-18$) is tested using a model trained on the data from the other two lakes from both winters.}%
\vspace{-0.5em}
\begin{tabular}{|c|c|c|c|c|}
\hline
\diagbox[]{IoU}{Lake} & Sils & Silvaplana  & St.~Moritz \\ \hline
Non-frozen & 96.7\% & 93.3\% & 85.6\% \\
%\hline
Frozen & 96.4\% & 92.7\% & 82.9\% \\
\hline
\textbf{Mean} & 96.5\% & 93.1\%  & 84.3\% \\
\hline
\end{tabular}
\label{tab:iou_lakes} 
\end{table}
\normalsize

\par
\begin{figure*}
    \centering
    \subfigure[Time series of percentage of non-frozen pixels for lake Silvaplana from winter $2017$-$18$.]{\includegraphics[width=0.82\textwidth]{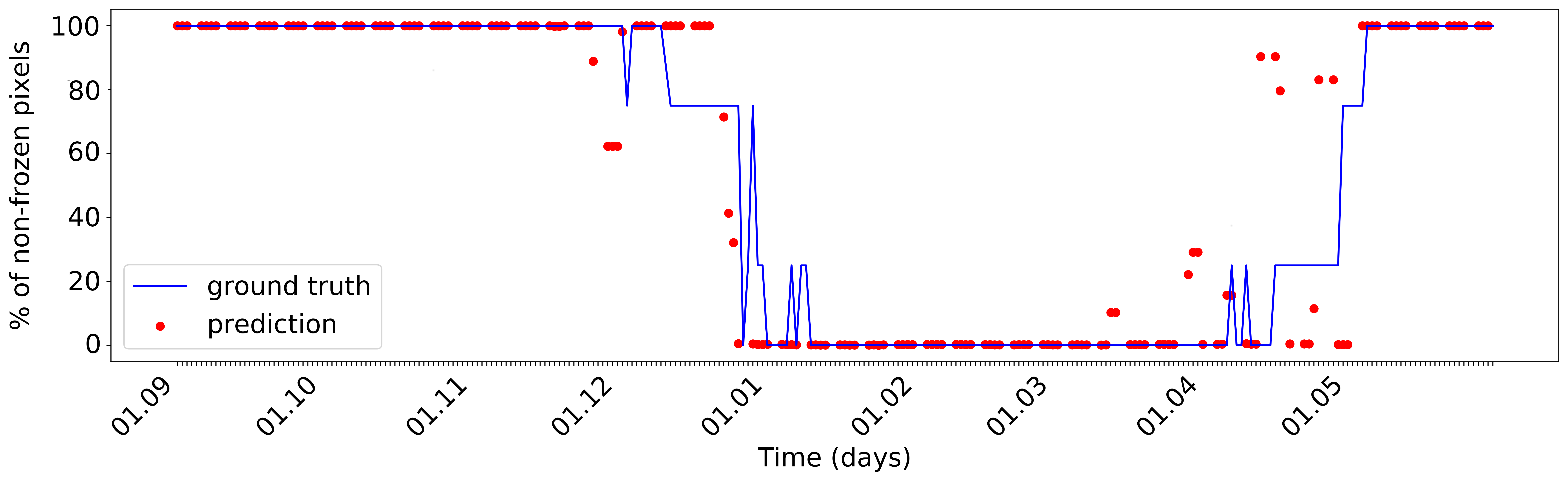}}\\
    \vspace{-1em}
    \subfigure[Temperature (temporal moving average of daily median with window size of $7$ days) from the nearest meteo station.]{\includegraphics[width=0.81\textwidth]{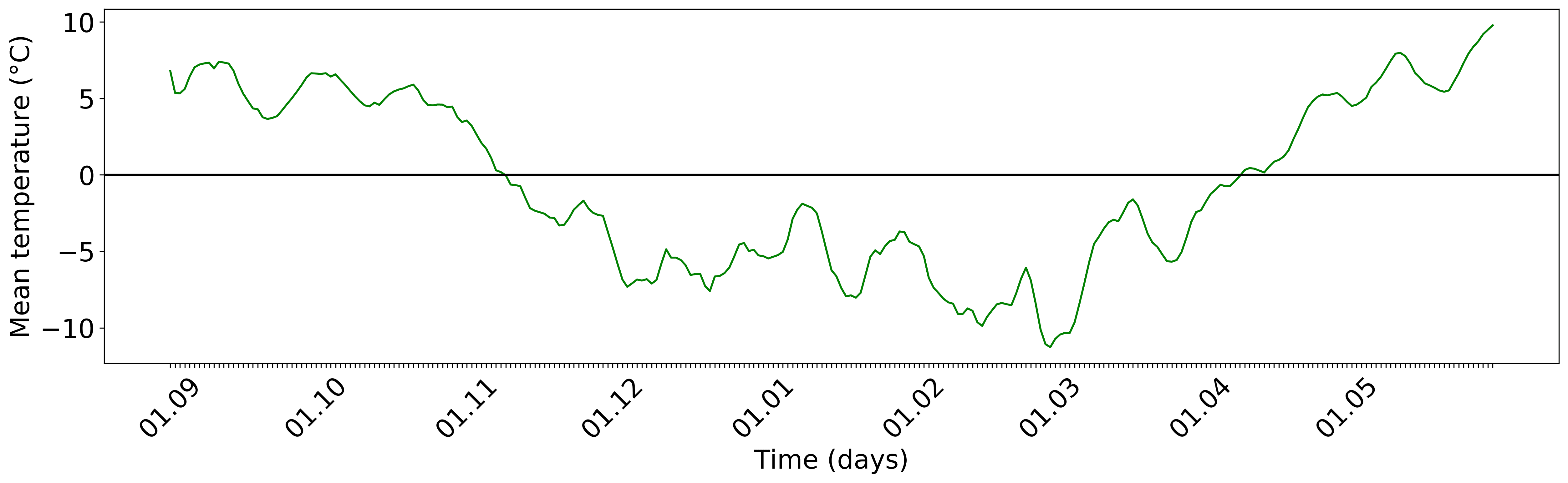}}
    \vspace{-1em}
    \caption{Correlation of our results (winter $2017$-$18$) on lake Silvaplana with the ground truth and the auxiliary temperature data. Best if viewed on screen.} 
\label{fig:time_series}
\end{figure*}

\textbf{Quantitative results: Time-series.}  The \textit{ice-on} and
\textit{ice-off} dates are of particular interest for climate
monitoring. From the per-day semantic segmentation results, we
estimate the daily percentage of frozen surface for each observed
lake. Thus, for each available SAR image, we compute the percentage of
frozen pixels, throughout the entire winter. An example time series
for lake Silvaplana in winter $2017$-$18$ is shown in
Fig.~\ref{fig:time_series}a. Although we do not have per-pixel ground
truth on the partially frozen transition days, we know whether the
lake has more water (shown with a value of $75\%$ in the ground truth)
or more ice/snow (shown with a value of $25\%$ in the ground
truth). Even though some miss-classifications exist during the
transition days, the non-transition days are almost always predicted
correctly, likely because the network was trained solely on
non-transition days. For a comparison, we also plot the time series of
temperature values (sliding window mean of the daily median,
window size $7$ days) obtained from the nearest meteo station in
Fig.~\ref{fig:time_series}b. Sub-zero values in this graph correlate
(with some time lag) with the period in which the lakes are fully or
partially frozen.

\begin{figure*}[]
  \centering
  \includegraphics[width=.95\linewidth]{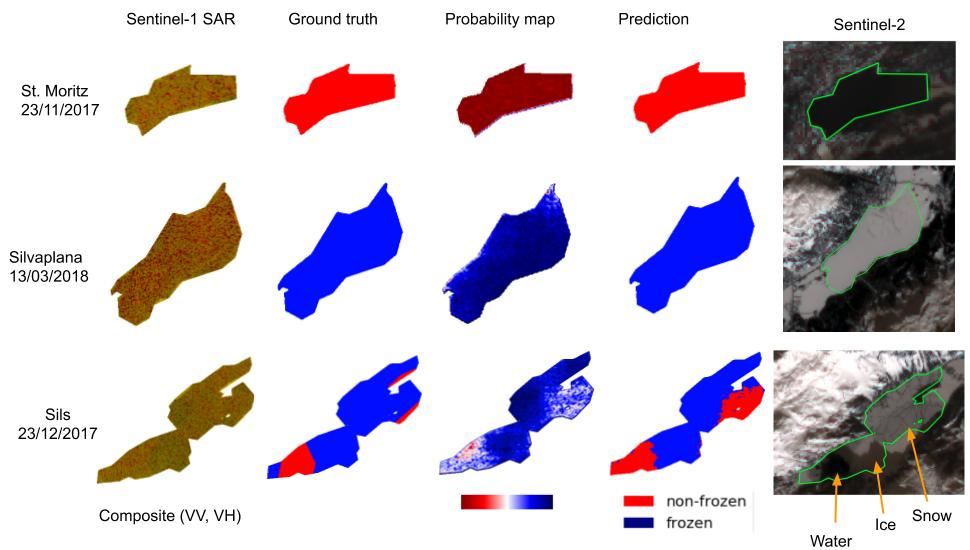}
\caption{Qualitative results for lake St.~Moritz
  on a non-frozen day (row $1$), lake Silvaplana on a frozen day (row
  $2$), and lake Sils on a transition day (row $3$). For each lake we show
  the Sentinel-1 composite image (column $1$), the ground truth (column
  $2$), the predicted frozen probability map derived from the network logits (column $3$), and the
  corresponding binary classification map (column $4$). Additionally, column $5$ shows
  a corresponding Sentinel-2 image for better visual
  interpretation.}
  \vspace{-1.2em}
\label{fig:qual_analysis}
\end{figure*}
\par
\textbf{Qualitative analysis.}\label{sec:qualitative} Exemplary
qualitative results are depicted in Fig.~\ref{fig:qual_analysis}. We
show the classification results on frozen, non-frozen, and transition
dates along with the probability map (blue means higher probability of
frozen, red means higher probability of non-frozen). For better
interpretation of the result, especially for the transition date, we
show the corresponding image from Sentinel-2.

\begin{table*}[]
    \small
    \caption{Per-class- and mean IoU values of frozen and non-frozen classes with different polarisations (left table) and orbits (right table). Data from all three lakes from winter $2016$-$17$ was tested using a model trained on the data from all three lakes from winter $2017$-$18$. \textit{Asc} and \textit{Dsc} denotes ascending and descending orbits respectively.}
    \vspace{-0.5em}
    \begin{minipage}{.5\linewidth}
    %\caption{A}
    %\hspace{-2em}
    \centering
     \begin{tabular}{|c|c|c|c|c|c|}
        \hline
        \diagbox[width=8em]{IoU}{Polarisation} & VV, VH & VH  & VV & VV, VV\_prev  \\ \hline
        Non-frozen & 91.0\% & 71.0\% & 87.7\% & 93.3\% \\
        %\hline
        Frozen & 90.6\% & 60.1\% & 86.6\% & 91.9\% \\
        \hline
        \textbf{Mean} & 90.8\% & 65.9\%  & 87.1\% & 92.6\%\\
        \hline
    \end{tabular}
    \end{minipage}%
    \begin{minipage}{.5\linewidth}
      \centering
        %\caption{B}
        %\hspace{-1.5em}
        \begin{tabular}{|c|c|c|c|c|}
        \hline
        \diagbox[]{IoU}{Direction} &  Asc, Dsc  & Dsc & Asc \\ \hline
        Non-frozen & 91.0\% & 84.7\% & 85.9\% \\
        %\hline
        Frozen & 90.6\% & 82.7\% & 84.3\% \\
        \hline
        \textbf{Mean} & 90.8\% & 83.7\%  & 85.1\% \\
        \hline
    \end{tabular}
    \end{minipage}
    \label{tab:iou_polarize_flightDirection} 
\end{table*}
\normalsize

\textbf{Miscellaneous experiments.}\label{sec:miscellaneous} In all
the experiments reported so far, we used the data from all four orbits
(both ascending and descending) and both polarisations (VV and VH). To
study the individual effect of polarisations VV and VH, we drop either
of them and report the corresponding results on Table
\ref{tab:iou_polarize_flightDirection} (left). Note that mIoU drops by
almost $25\%$ when VV is left out, while it drops by only $3.7\%$
without VH, confirming the significance of polarisation VV for lake
ice detection. This finding also aligns with the visual
differences in Fig.~\ref{fig:distribution}. However, while VH appears
to be less discriminative overall, it is much less affected by wind speed --
see Fig.~\ref{fig:dist_wind}. We believe that using also VH may
improve robustness in windy conditions, where discriminating water
from ice/snow should be particularly challenging because of increased
surface roughness due to waves. However, we do not have enough days
with strong wind to quantitatively corroborate this hypothesis.  From
our current data it appears that the system can handle calm and
moderately windy days practically equally well. 

\begin{figure}[]
    \begin{center}
	    \includegraphics[width=0.65\columnwidth]{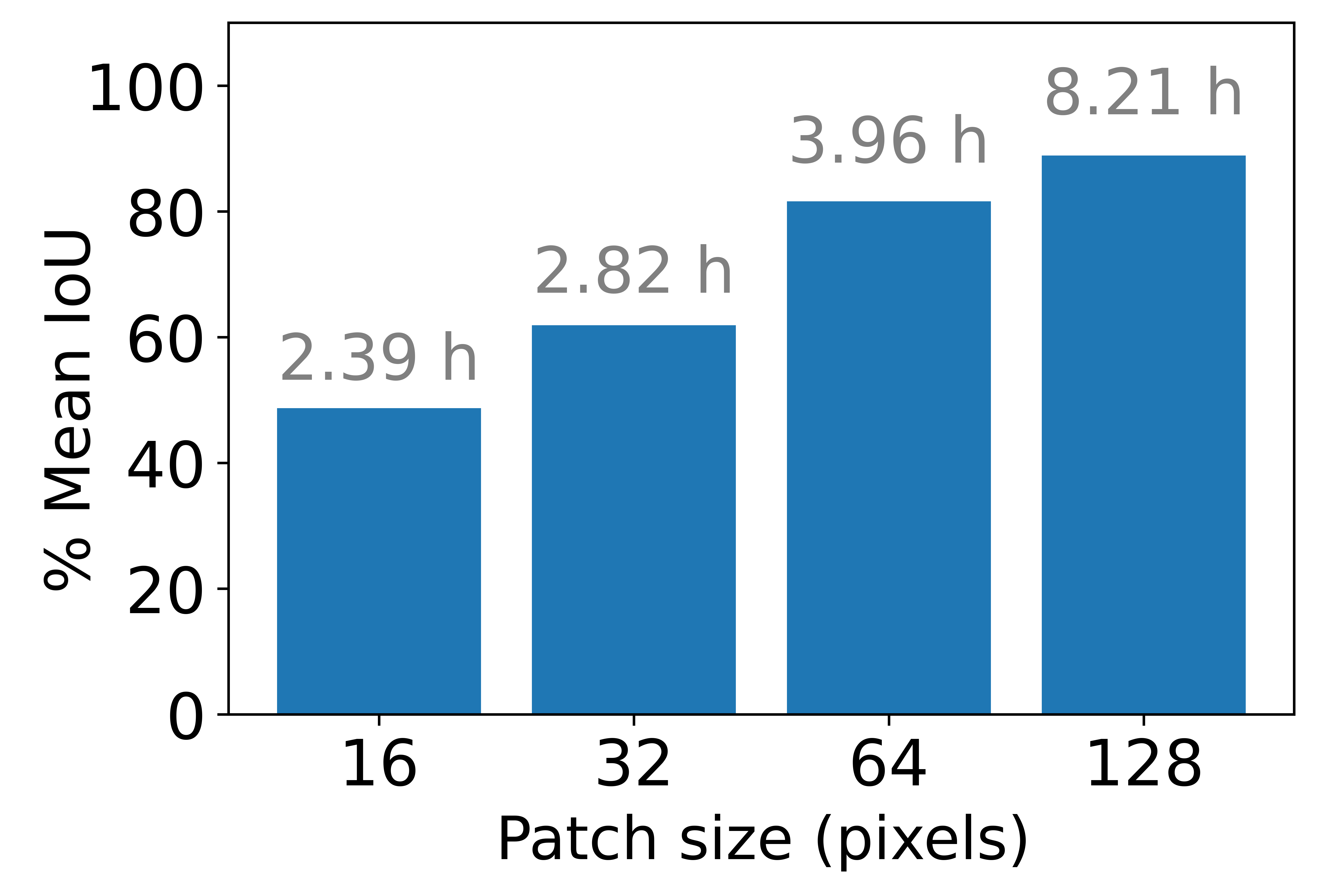}
	    \vspace{-1.5em}
	\caption{Mean IoU values obtained with different input patch
          sizes. The total time taken to complete training and testing
          is also indicated, in hours (h). The results are for all
          three lakes combined, training on $2016$-$17$ and testing on
          $2017$-$18$.}
    \label{fig:patch_size}
    \end{center}
\end{figure}

In another experiment, we drop VH but use VV from two temporally adjacent acquisitions (VV and \textit{VV\_prev}), thus simultaneously feeding the network with data from two different days. The mIoU rises by ~$2\%$, see Table \ref{tab:iou_polarize_flightDirection} (left). However, we noted some stability issues especially during fully frozen days. We believe this is due to the fact we removed VH. We did another experiment to check the effect of acquisition time. Here, we used the data from both VV and VH. However, we drop the data from some orbits (see Table \ref{tab:orbits}). 
%The satellites scan the Region Sils around $5:10$ pm in ascending orbits ($15$ and $117$), respectively around $5:30$ am in descending orbits.
Table \ref{tab:iou_polarize_flightDirection} (right) shows that the
mIoU drops by $7.1\%$ and $5.7\%$ respectively when the data from only descending
orbits ($66$ and $168$) or only ascending orbits ($15$ and $117$) were
used. A final experiment was done to assess the influence of the (rectangular) training patch size, see Fig. \ref{fig:patch_size}. Somewhat surprisingly, even the move from
an already large context of $64\times64$ pixels to $128\times128$
pixels ($1.3\times 1.3$ km) still brought a marked improvement, hence
we always use that patch size in our system.

\vspace{-0.5em}
\section{CONCLUSION AND OUTLOOK}
We have described a system for reliable monitoring of lake ice based
on Sentinel-1 SAR imagery, with the potential to retrieve long,
consistent time series over many years (assuming continuity of the
satellite mission).
The proposed method has been demonstrated for three different Swiss
lakes over two complete winters, and obtains good results
(mIoU 90\% on  average, and \textgreater84\% even for the most difficult lake), even when generalising to an unseen winter
or lake.
Given the main advantage of SAR data for our purposes -- its ability
to observe with very good spatial and temporal resolution independent
of clouds -- we see the possibility to extend our method into an
operational monitoring system.
A logical next step would be to process longer time series, which
unfortunately is not yet possible with Sentinel-1. It is quite
possible that even a moderate time span, say 20 years, would suffice
to reveal trends in lake freezing patterns and perhaps also
correlations with climate change.
Another future direction is an integrated monitoring concept, using SAR together with optical satellite imagery and optionally images
from webcams, to ensure reliable
identification of \textit{ice-on} and \textit{ice-off} dates within
the GCOS specification of $\pm 2$ days.
%\input{chapters/section7_appendices}

% Acknowledgments should only be added after acceptance
%
%\vspace{-0.75em}
%\section*{ACKNOWLEDGEMENTS}\label{ACKNOWLEDGEMENTS}
\par
\textbf{ACKNOWLEDGEMENTS.} This work is part of the project \textit{Integrated lake ice monitoring and generation of sustainable, reliable, long time series} funded by Swiss Federal Office of Meteorology and Climatology MeteoSwiss in the framework of GCOS Switzerland. Also, we thank Anton B. Ivanov from Skoltech for his support.
\vspace{-1.0em}
{
	\begin{spacing}{1.17}
		\normalsize
		\bibliography{authors} % Include your own bibliography (*.bib), style is given in isprs.cls
	\end{spacing}
}
%\section*{APPENDIX (Optional)}\label{APPENDIX}
%Any additional supporting data may be appended, provided the paper does not exceed the limits given above. 

%\vspace{1cm}
%\textit{Revised January 2020}
\end{document}